\documentclass[a4paper,twoside]{article}
\newcommand{\affil}[1]{$^{\rm #1}$}
\usepackage[authoryear]{natbib}
\bibpunct{(}{)}{;}{a}{}{,}
\usepackage{graphicx}
\usepackage{amssymb}
\date{} 
\title{\large\bf\flushleft Remarkable symmetries in the Milky Way disc's magnetic field}
\author{\parbox{\textwidth}{\flushleft
\vspace{-0.5cm}
{\it P. P. Kronberg\affil{A,B,E}, K.~J. Newton-McGee\affil{C,D}}\\
\vspace{0.4cm}
{\small \affil{A}\,Dept. of Physics, University of Toronto, Toronto M5S 1A7, Canada}\\
{\small \affil{B}\,IGPP, Los Alamos National Laboratory, M.S. T006, Los Alamos NM 87545, USA}\\
{\small \affil{C}\,Sydney Institute for Astronomy, School of Physics, The University of Sydney, NSW 2006, Australia}\\
{\small \affil{D}\,Australia National Telescope Facility, CSIRO, PO Box 76, Epping NSW 1710, Australia}\\
{\small \affil{E}\,Email: kronberg@physics.utoronto.ca}}}
\begin{document}
\twocolumn[
\begin{changemargin}{.8cm}{.5cm}
\begin{minipage}{.9\textwidth}
\vspace{-1cm}
\maketitle

\small{\bf Abstract:} 

Using a new, expanded compilation of extragalactic source Faraday rotation measures (RM) we investigate the broad underlying magnetic structure of the Galactic disk at latitudes $|b|$ $\lesssim 15^{\circ}$ over all longitudes $l$, where our total number of RM's in this low-latitude range of the Galactic sky is comparable to those in the combined Canadian Galactic Plane Survey(CGPS) at $|b| < 4^{\circ}$ and the Southern Galactic Plane (SGPS) $|b| < 1.5^{\circ}$ survey. We report newly revealed, remarkably coherent patterns of RM at $|b|$ $\lesssim 15^{\circ}$ from $l \sim 270^{\circ}$ to $\sim 90^\circ$ and RM($l$) features of unprecedented clarity that replicate in $l$ with opposite sign on opposite sides of the Galactic center. They confirm a highly patterned bisymmetric field structure toward the inner disc, an axisymmetic pattern toward the outer disc, and a very close coupling between the CGPS/SGPS RM's at $|b| \lesssim 3^{\circ}$ (``mid-plane'') and our new RM's up to $|b| \sim 15^{\circ}$ (``near-plane'').

Our analysis also shows the approximate $z$-height -- the vertical height of the coherent component of the disc field above the Galactic disc's mid-plane -- to be $\sim 1.5$kpc out to $\sim 6$ kpc from the Sun. This identifies the approximate height of the transition layer to the halo field structure. We find no RM sign change across the plane within $|b| \sim 15^{\circ}$ in any longitude range. The prevailing {\it disc} field pattern, and its striking degree of large scale ordering confirm that our side of the Milky Way has a very organized underlying magnetic structure, for which the inward spiral pitch angle is $5.5^{\circ}\, \pm 1^{\circ}$ at all $|b|$ up to $\sim 12^{\circ}$ in the inner semicircle of Galactic longitudes. It decreases to $\sim 0^{\circ}$ toward the anticentre.

\medskip{\bf Keywords:} magnetic fields -- methods: observational -- Galaxy: disc -- Galaxy: structure -- radio continuum: ISM

\medskip
\medskip
\end{minipage}
\end{changemargin}
]
\small

\section{Introduction}

Early extragalactic source (e.g.r.s) RM probes of the Milky Way magnetic field by Gardner \& Davies (1966), and the first pulsar RM analyses by Manchester (1972, 1974) revealed a prevailing magnetic field in the local spiral arm  that is clock-wise as viewed from the north Galactic pole. Further study with a 555-source extragalactic RM dataset produced the discovery of a large scale field reversal in the Sagittarius-Carina arm (Simard-Normandin \& Kronberg,  1979). Subsequent numerical analyses by Simard-Normandin \& Kronberg (1980) confirmed a prevailing local field pointing to $l \sim 75^\circ$, a more complex pattern in the north Galactic hemisphere, an anomalously high off-plane, high  RM zone (``Region A''), a $\sim 30^\circ$ angular autocorrelation scale of RM's off the plane, and a Galactic magneto-ionic scale height of $\sim$ 1.8 kpc. 

Confirmation of a large scale Sagittarius-Carina field reversal has since come from several e.g.r.s and pulsar RM studies  (Rand \& Lyne, 1994; Han {\it et al.}, 1999; Frick {\it et al.}, 2001 \& Weisberg {\it et al.}, 2004). Variants of this model have been proposed by Han {\it et al.}, (2006) $-$ a bisymmetric model with field reversals at each arm/interarm interface; Vall\'{e}e (2008) $-$ a concentric ring model having one field reversal, and Sun {\it et al.}, (2008) who concluded that a combined axisymmetric and toroidal model best describes the Galactic magnetic field. Men {\it et al.}, (2008), using pulsar RM's, concluded that neither a concentric ring model nor a bi-symmetric or axi-symmetric field model alone, suffices when attempting to describe the Galactic magnetic field.

The high density of e.g.r.s. RM's in the $|b| < 4^\circ$ Canadian Galactic Plane Survey (CGPS) (Brown, Taylor, \& Jackel, 2003), 
and in the $|b| < 1.5^\circ$ Southern Galactic Plane Survey (SGPS)  (Brown {\it et al.}, 2007) were used to probe both small-scale 
(Haverkorn {\it et al.}, 2006), and larger-scale field structures (Brown {\it et al.}, 2007). The best fit model of the latter has 
a clockwise field everywhere except for two reversals, one in the Scutum-Crux arm and the other in the molecular ring at a smaller 
Galactocentric radius of $\approx$ 3.7 kpc. Since this work was completed, Van Eck et al. (2011) extended the previous results of 
Brown {\it et al.} (2003, 2007) above after closing observational gaps in the ``mid plane'' ($|b| \lesssim 3^{\circ}$ sources. Also since this work was completed Taylor {\it et al.} (2009) published a very large RM compilation using 2 closely spaced L-band 
wavelengths in the NVSS survey. However the NVSS catalogue's zone of avoidance in the Southern celestial sphere gives it limited applicability for this investigation since it removes {\it ca.} 1/3 of the of 
$0^{\circ} < |b| < 15^{\circ}$ ``near plane'' zone RM's which is the focus this paper. 

Conflicting conclusions among some of the above publications illustrate the difficulties in unravelling 
the magnetic structure of the Galaxy. This paper presents an all-sky plot of the Milky Way's smoothed RM pattern, but focuses solely on the global magnetic structure in the near-plane zone up to $|b| \sim 15^{\circ}$. 
It uses an improved and expanded set of extragalactic source Faraday RM's of unprecedented average accuracy, most of which were derived 
from linear polarization measurements over a wide range of wavelengths, from $\sim \lambda 2$cm to $\sim \lambda 23$cm. 

The smoothing ``beam'' of 
$\sim 21^{\circ}$ is comparable to a spiral arm (and inter-arm) width  dimension, thereby providing a resolution comparable to some current 
magnetic field images of nearby external galaxies such as M51. In the ``above disk'' RM sky, the smoothed RM
patterns are quite different from those near the disk. Interpretation of the $|b| \gtrsim 12^{\circ}$ RM sky requires a 3-D modeling analysis, and this will  be the subject of a following paper.        

\section[]{The data, methods and analysis}

We use a new 2257-RM compilation of $\simeq$ 1500 revised and more accurate e.g.r.s. RM's of our own (Newton-McGee \& Kronberg, in prep.), combined with the published CGPS/SGPS RM's, and smaller published lists from Klein {\it et al.} (2003), and from Mao {\it et al.} (2008).  These RM surveys also enable us to explore the match between the ``in-plane'' RM's, mostly the CGPS/SPGS data, and the ``near-plane'' RM's from our own new data, both above and below the Galactic plane.

Figure 1 shows an equal area, all-sky projection of the smoothed RM's. The RM smoothing method is described in Simard-Normandin \& Kronberg (1980), in which an iterative calculation of the mean RM was performed at each source location, RM($l_i$,$b_i$), using all neighbour source RMs within a 15$^\circ$ radius.
At each iteration for a given source, RM values $>$ 1.3$\sigma$ from the mean were eliminated as ``outliers'', 
and a minimum of 4 RM's was required to define a smoothed RM value at that ($l$,$b$).
We retested their outlier rejection criterion of $\sim 1.3\sigma$, using both higher and lower $\sigma $ trial rejection levels, and confirm that it is optimal. The criterion was to minimize the {\it distribution width} in $\left | RM \right |$ 
of Galaxy-corrected RMs using only sources at high $|b|$ where the Galactic contribution to RM is smallest compared to the source-to-source RM scatter. 

In the Southern Galactic plane from $l \approx 270^\circ$ to $l \approx 355^\circ$  one major RM sign change occurs at our smoothing resolution, at $l \approx$ 310$^\circ$ (Figure 2(a)). We note that smaller scale changes of RM sign on smaller scales have been found in the ``mid-plane'' zone by  Brown {\it et al.} (2007), and Van Eck et al.(2011). Our smaller number of sign changes is purely an effect of our smoothing half-width, $\approx 21^\circ$, which averages over some (real) smaller scale RM sign changes. This also applies to other $l$-zones near $b$ = 0$^\circ$, and at some higher latitudes that we do not discuss in this paper. Our method effectively isolates larger scale RM features in the vicinity of the Milky Way disc -- as they would be seen by an extragalactic observer. 

The analysis that follows focuses on $|b|$ $\lesssim$ $15^\circ$. It imposes cuts of the RM data variously, and as discussed below, 
at $|b| = 2^{\circ}$, $5^{\circ}$, $10^{\circ}$ $12^{\circ}$,$--$ and finally $20^{\circ}$ to explore the outer $|b|$ limits of the 
disc-associated field ($\S 3.3$). Putting these angles in the context of disc $z$-heights, an RM 
sightline at $|b| = 10^{\circ}$ corresponds to a 1.4 kpc $z$-height in the Milky Way disk at 8 kpc from the Sun. 
This reference value of 1.4 kpc is similar to several estimates of the disc magneto-ionic scale height (e.g. Simard-Normandin \& Kronberg (1980) and more recent
analyses). It is also comparable to the galaxy's synchrotron radiation thick disc height of $\sim 1.2 - 1.9$ kpc (Beuermann et al. 1985), and 
to the ionized gas $z$-height estimate (Gaensler et al. 2008).
The following analysis indicates clear large$-$scale 
patterns and symmetries on our side of the Milky Way disc that were either not previously apparent, or as clear.

\begin{figure*}
\centering
\begin{minipage}{145mm}
\includegraphics[width=16cm]{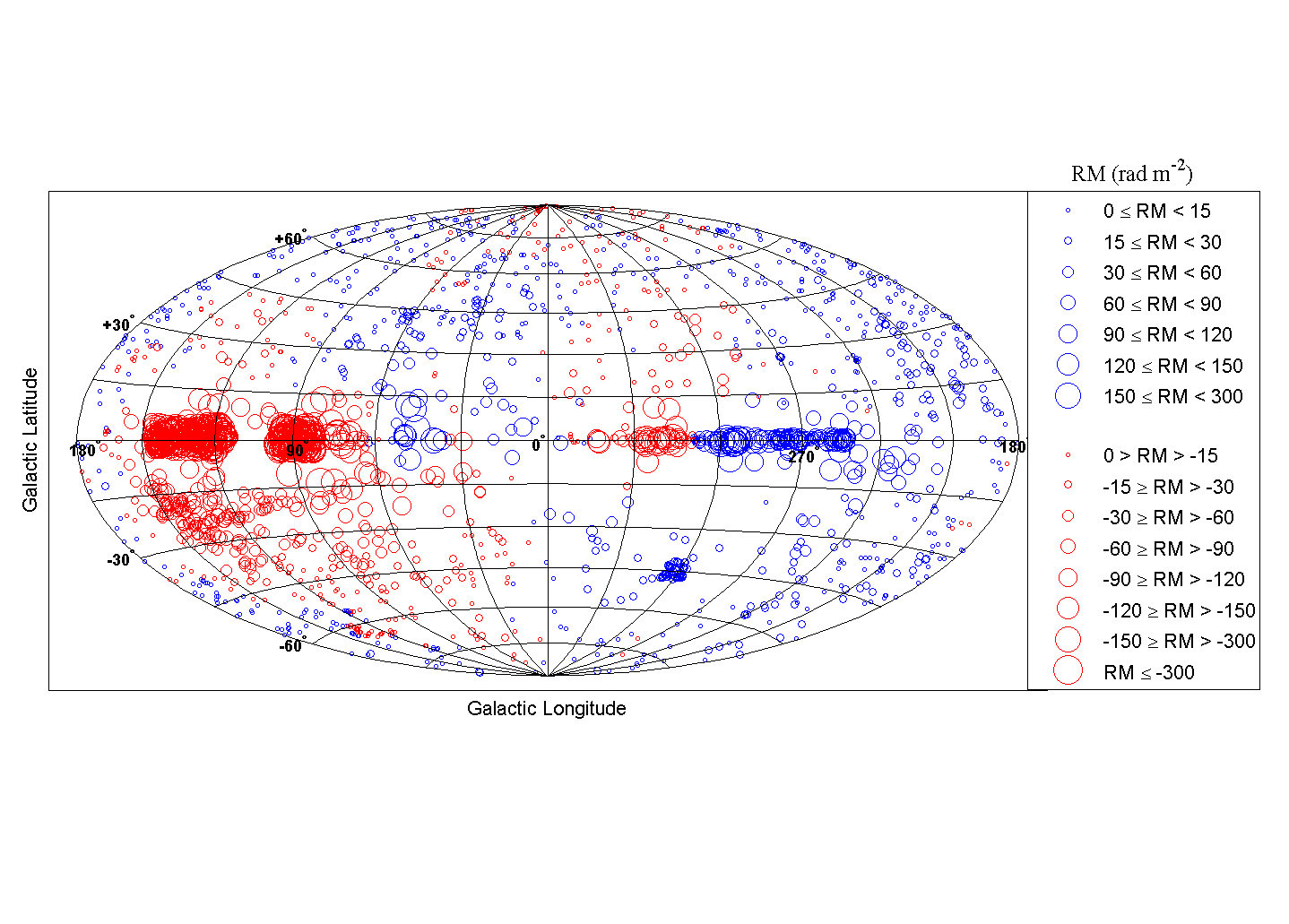}
\caption[Allsky RM plot]{An all-sky equal-area projection RM plot of the smoothed RM's 
from our 2257-source compilation
of extragalactic source RM's. The smoothing method is described in the text.}
\end{minipage}
\end{figure*}

\section[]{Highlights of the disc magnetic field analysis}

\subsection[] {Determination of the prevailing magnetic field direction on our side of the Galaxy}

The smoothed RM's for $|b| < 10^\circ$ are shown in Figure 2(a). The second Galactic $l$-semicircle is then folded, at $l = 0^\circ$, onto the first. When we then reverse the RM signs in this folded semi-circle (Figure 2(b)) a clear symmetrical/ anti-symmetrical pattern emerges in the smoothed RM's. Specifically, the {\it forms} of the smoothed $/$averaged RM's are strikingly similar but they have opposite signs on opposite sides of the Galactic center. 

Figure 2(b) also shows clearly that the Galactic centre direction, $l$ = 0$^\circ$ is not the correct symmetry axis: An 11$^\circ$ relative shift of the folded datasets needs to be applied $-$ {\it i.e.} fold, invert sign, {\it and} shift, in that order. This same order is followed throughout the paper. For optimum determination of this shift, we examined RM's between $44^\circ$ to $76^\circ$ of the post-shift folding centre, and applied a least squares minimization of the residual as a function of shift angle. Figure 2(c) shows the remarkably good overlap that results. This shift corresponds to a pointing of the prevailing local B-field to $l = 84.5 \pm 1^\circ$, {\it i.e.} inward from the $l = 90^\circ$ tangential by 5.5$\pm 1^\circ$. 

Independently, when we examine the all-sky RM map in Figure 1 at higher latitudes near $l$ = 180$^{\circ}$, a similar 
change from negative to positive RM's is evident near the anti-centre direction nearly all the way to the South Galactic Pole. Above the plane, a similar negative-to-positive RM ``border'' can be traced as far as $b \simeq + 45^{\circ}$, where it then follows a different sign boundary near to the $b = 45^{\circ}$ line.
We do not attempt to interpret this latter northern high latitude wandering here, except to note that Frick et al. (2001) attribute the north-south pattern asymmetry at higher $|b|$ to either a stronger field in the southern Galactic hemisphere or by the Sun being located close to the top of a magnetic loop.

The $l = 84.5 \pm 1^\circ$ magnetic direction is very close to the value of $l = 82.8 \pm 4.1^\circ$ for local field
lines derived from starlight polarizations by Heiles(1996), after correcting for larger scale spiral arm curvature. In general, any magnetic pitch angle determination will slightly depend on precisely what distance range from the
Sun is defined as local. Since our Galactic disk RM's represent typical pathlengths up to several kpc through the 
disc (see $\S 2$), this remarkably close agreement between the present analysis and the Galactic curvature-corrected starlight polarization results, 
may be partly coincidental. The agreement with completely independent starlight polarization results is nonetheless worth
noting.  

The patterns seen in Figure 2 indicate that an extragalactic observer would see at least our side of the Milky Way disc 
with a very patterned and organized magnetic spiral structure, with a local pitch angle of 5.5$^\circ$ near the location of the Sun. This, combined with the symmetrical/antisymmetrical RM patterns seen in Figure 2 strongly indicate that that the Milky Way disc is laid out like other ``magnetically
organized`` spiral galaxies.

\subsection[] {A test for ``magnetic coupling'' between the mid-plane disc and the immediate ``above plane'' zones}

We now repeat the procedures in Figure 2, using sources having $|b| < 10^\circ$ but omit the RM's at the very low latitudes ($|b| < 2^\circ$), {\it i.e.} close to the Galactic mid-plane. We also excluded any CGPS RM's that are found up to $|b| = 4^\circ$. This leaves 210 RM's from our own compilation at $2^\circ < |b| < 10^\circ $, and these are spread around the entire $l$ circle. As mentioned above, all the ($l$,$b$) subset exclusions were done {\it before} re-smoothing.

Figure 3 shows patterns that are virtually identical with Figure 2(a), including the crossover points.
Application of our fold-reverse-shift-optimize technique in Figure 2(b) and (c) yields the same best fit spiral pitch angle, 
$l_{0} = 84.5^\circ \pm 1^\circ $ as when the $|b| < 2^\circ  $RM's are included. 
The same result (not shown) is also obtained for the ``$|b| < 2^\circ$ only'' subset of RM's. This result is notable for two reasons: 
(1) The close agreement occurs between two key non-overlapping datasets $-$ our source sample above $|b| = 2^{\circ}$, and the 
CGPS/SGPS RM's at the lowest $|b|$'s. This also provides a convincing posterior check on the reliability of both RM data sets. 
(2) Astrophysically, it demonstrates a close coupling between the magnetic field structure in the ISM of the in-plane zone, 
and at $|z|$ heights up to $\sim$ 1 kpc away from the mid-plane.
 
In the above we did not distinguish RM data above and below the plane. We next test for magnetic sign asymmetries/antisymmetries {\it across} the Galactic plane. As in Figure 3, we omit all RM's at $|b| < 2^\circ $. To have adequate statistics for this purpose, we slightly extend the upper boundaries in $|b|$ to $\pm 12^\circ $. Then we re-apply the fold-reverse-shift-optimize technique separately, for RM's at $-12^\circ < b < -2^\circ $ and $+12^\circ > b > +2^\circ $. 

The same best fit procedure for the prevailing Galactic B-field direction gives $l_{0} = 84.5 \pm 2^\circ $ in the negative $b$ range, and $l_{0} = 86.5 \pm 2^\circ $ in the positive $b$ range. The magnetic field patterns in $l$ are therefore the same within the error limits, above, and below the plane, 84.5$^\circ $ (or 11$^\circ $relative shift after folding). They also agree with the ``full-plane'' results in Figure 2.

The respective plots are combined in Figure 4. Here we see that, at our $\sim 21^\circ$ smoothing resolution, the sign of the 
prevailing disc magnetic field pattern does not change through the mid-plane. More explicitly, within the longitude boundaries 
$l \sim 280^\circ $ to $\sim 60^\circ $ (the ``inner'' longitude range) where the RM signs change with $l$, these changes 
replicate locally with even symmetry {\it i.e.} RM($+b$) = RM($-b$). 

It is important here to emphasise that, at these inner longitudes, the even RM sign symmetry across the mid-plane just demonstrated in Figure 4, holds only at latitudes within the  $|b| \lesssim 15^{\circ}$ zone of the galactic disk that that we have been discussing. At some ``inner'' longitudes
sign changes in certain $l$ ranges do occur {\it beyond} $|b| \sim 15^{\circ}$.  This can be seen in Figure 1, and partially seen in the (un-blanked portion of the) RM distribution of Taylor et al. (2009). The nearest approximation to our result on cross-plane sign symmetry result was obtained by Frick et al. (2001) in their wavelet analysis of the 555 RM dataset of Simard-Normandin et al.(1981). Our new combination of the CGPS and SGPS plus the $|b| \lesssim 15^{\circ}$ subset of our new  1500 RM's can now more precisely clarify how, and at what off-plane latitudes, systematic changes in RM sign with $b$ do occur. The RM sign symmetry in $b$ qualitatively confirms the conclusion of Frick et al. (2001), but contrasts with Jansson {\it et al.}'s (2009) conclusion of RM sign antisymmetry with respect to $b$. However RM antisymmetry does occur at some longitudes if all RM's at $|b| < 15^{\circ}$ are blanked out.

Outward of $l$ $\sim 280 ^\circ$ ($l < 280 ^\circ$) and of $l \gtrsim 60^\circ$, the Galactic field has the same prevailing sign above, in, and below the Galactic plane. This includes the ``in-plane'' RM regions of the CGPS and SGPS ($|b| < 4^\circ$ and $< 1.5^\circ$, respectively). 
Below the plane, the RM's keep the same prevailing sign at {\it all} $l$ ranges from $l \gtrsim$ 60$^{\circ}$ to $l \lesssim$ 330$^{\circ}$ right down to the South Galactic Pole. These results confirm that the ``{\it outer longitude}'' RM's symmetry with respect to the midplane is consistent with an overall quadrupole (even) symmetry of the large scale field. This question is further investigated in a following paper that analyses the entire $|b| > 10^{\circ}$ RM data of Figure 1. (Pshirkov et al. ${\it in~ prep.}$ 2011).  

Note the general point of distinguishing outer $l$'s ({\it e.g.} toward the anticentre) from true radial distance from the Sun to the outer 
Galaxy.

\begin{figure}
\centering
\includegraphics[width=8cm]{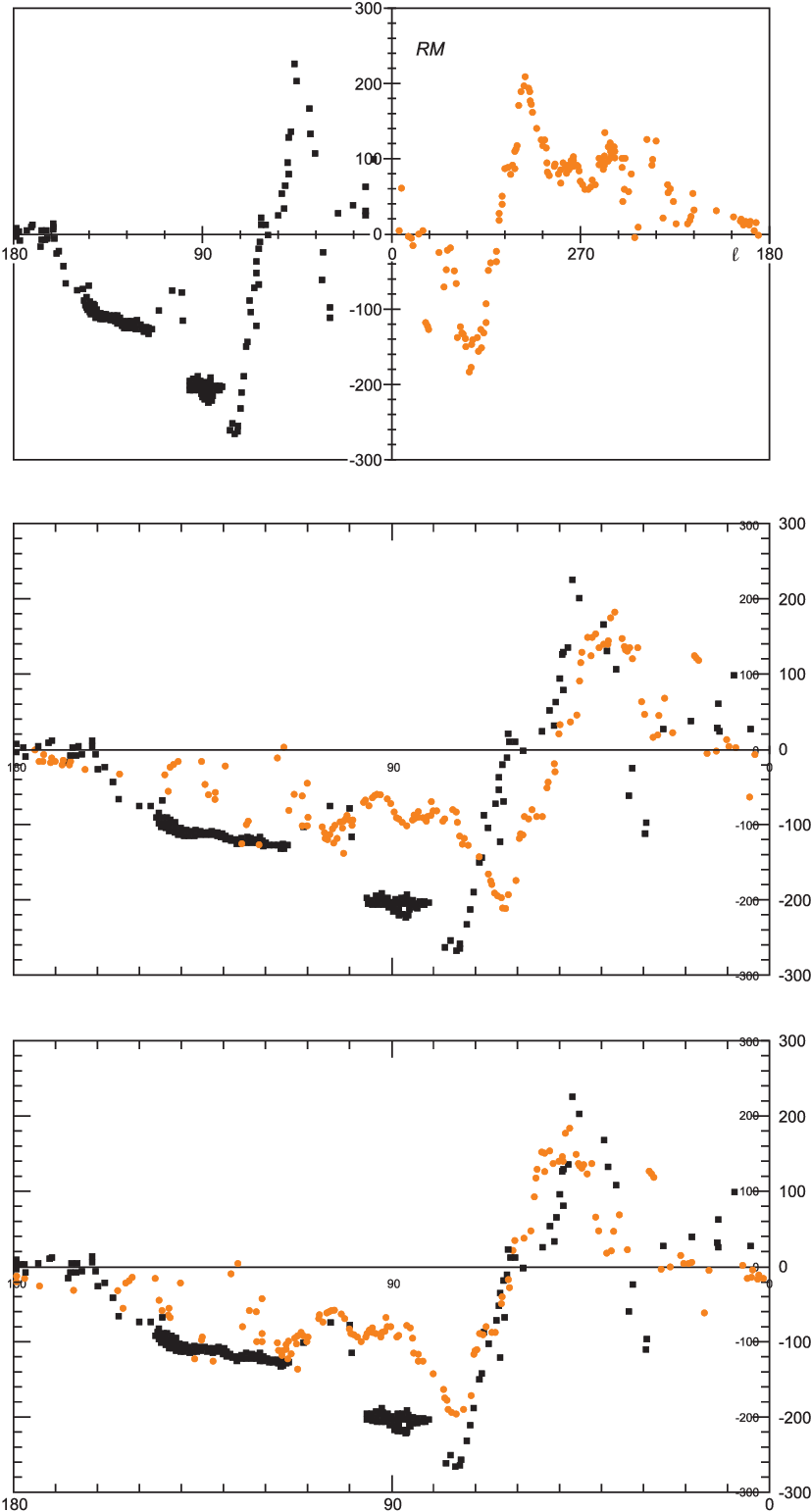}

\caption[The RMs as a function of longitude]{Panel (a) Plot of all smoothed RMs having $\left | b \right | < 10^{\circ}$  vs. $l$. 
Panel (b) shows the smoothed RM's with reversed sign, {\it and} ``flipped'' about $l$ = 0$^{\circ}$. Panel (c) is similar to (b) except 
that instead of the 180$^{\circ}$ rotation (flip) about $l = 0^{\circ}$, it is rotated about $l$ = 349$^{\circ}$(-11$^{\circ}$).}
 
\end{figure}

\begin{figure}
\centering
\includegraphics[width=8cm]{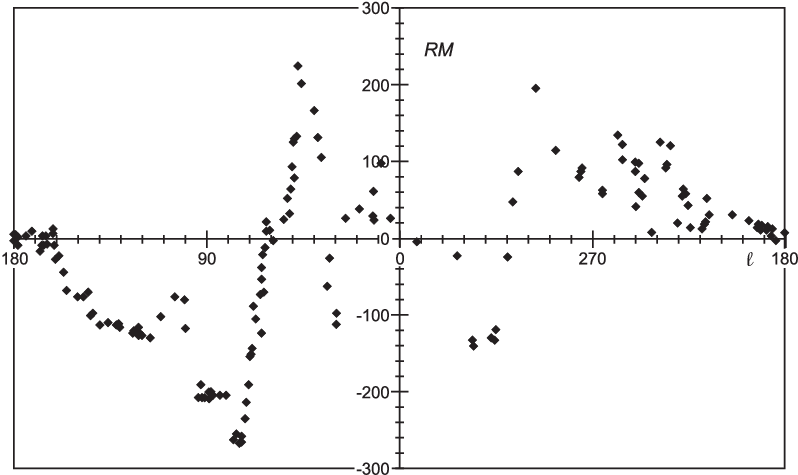}

\caption[Smoothed RM's excluding the CGPS and SGPS sources]{Plot of the smoothed RM's around the Galactic plane as in Figure 2(a), 
but excluding all the CGPS and SGPS sources, and all other sources at the lowest Galactic latitudes below $\left | b \right | = 2^\circ$.}
\end{figure}

\begin{figure}
\centering
\includegraphics[width=8cm]{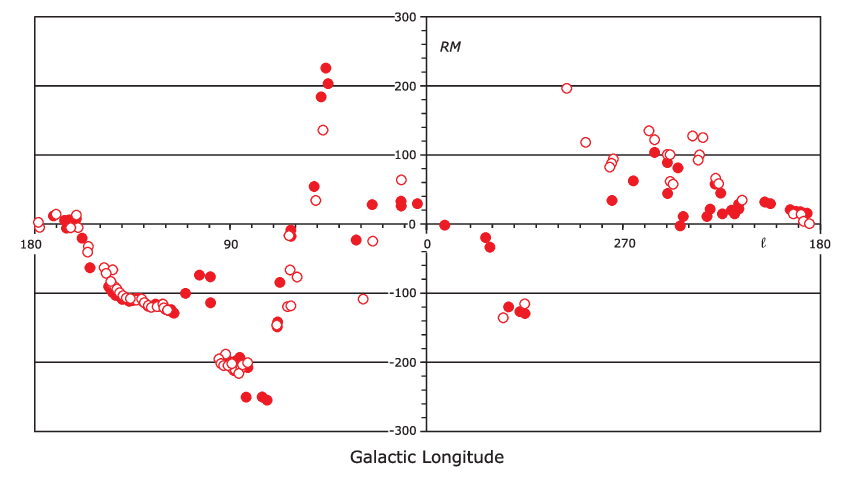}

\caption[Smoothed RM's separately above and below the Galactic plane]{Separate plots of the smoothed RM's around the Galactic plane, 
splitting by negative (open circles), and positive (filled circles) $b$, and extending the upper $|b|$ limit of data selection 
to $|12|^\circ $. It demonstrates the consistency of the prevailing magnetic field sign {\it across} the Galactic disc. }
\end{figure}

\begin{figure}
\centering
\includegraphics[width=8cm]{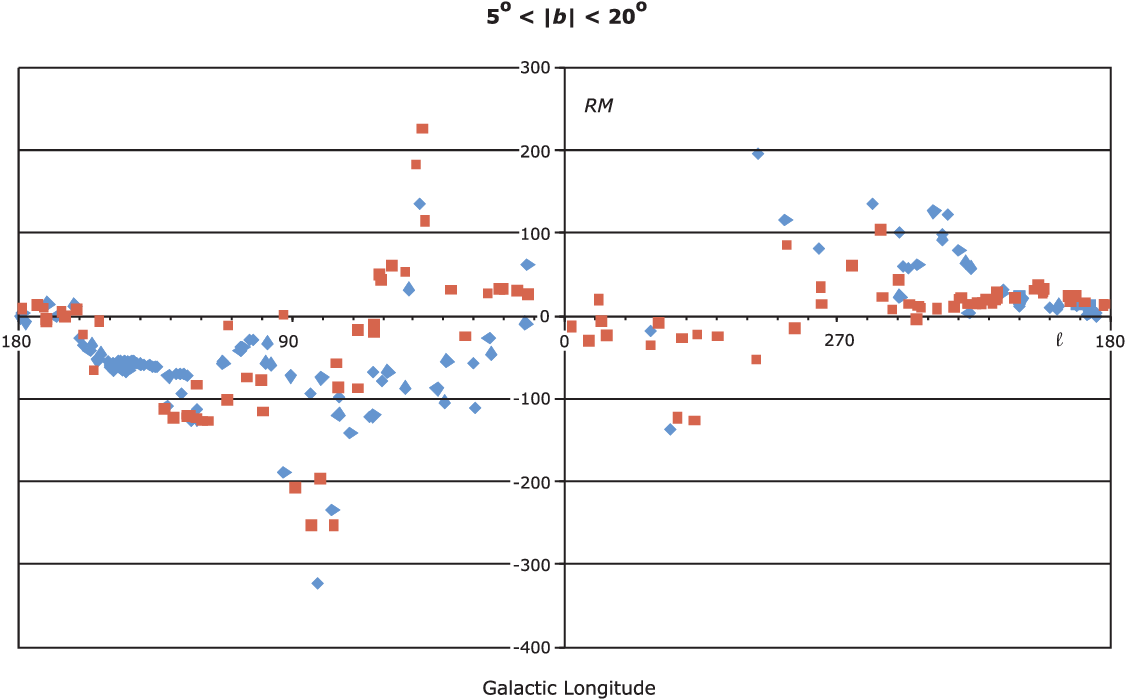}

\caption[Smoothed RM plot with raised latitude limits]{Plot of the smoothed RM's as in Figures 3a and 4, but including only RM 's
at a higher range of $20^\circ > |b| > 5^\circ $. Points above (diamond symbols) and below (square symbols) the Plane are distinguished.}
\end{figure}

\subsection[] {A new test for the $z$-extent of the regular disc field}

Finally, we revisit evidence for the thickness of the magnetic field coherence zone in our vicinity of the Milky Way disc. To do this, we raise the 
upper $|b|$ boundary further from the plane, recalling that $|b| = 10^{\circ} $ corresponds to a 1.4 kpc $z$-height at 8 kpc distance (approximately the distance to the Galactic centre). Our RM smoothing resolution at this distance is close to optimal for this purpose. To put our RM smoothing ``beam'' in a wider scale context: At 6 kpc distance our ``beam'' is roughly comparable to: (a) the spiral arm ionized gas full width, (b) a spiral arm width in H{\sc i} in the Galactic ($x$,$y$) plane, and also (c) approximately an inter-arm separation. 

As the selection boundaries of $|b|$ increase away from $|b| \sim 10^{\circ}$, the sharp features and symmetries seen in Figures 2 and 3 become progressively less distinct. Figure 5 shows all smoothed RM points having individual RM's pre-selected to $5^{\circ}$  $< |b| < 20^{\circ}$. While the distinctive shapes in Figure 2 and 3 can still be recognised, they are no longer as pronounced. That is, our fold-reverse-shift method no longer produces a clear merging of patterns, and the underlying spiral pattern ``fades away'' as we proceed toward the Galactic halo. In this way we have established the approximate transition zone between the Milky Way's magnetic disk and halo, which is in the range 1 - 2 kpc above the Galactic mid-plane. The location of this transition agrees with the conclusion of Jansson et al. (2009), based on a combination of the WMAP 22GHz data and a smaller set of RM's. 

\section[]{Summary} 
We have discussed new and striking underlying magnetic field patterns in the Galactic disc by suitably analysing an expanded and revised compilation of 2257 extragalactic RM's. The unprecedented clarity of our results comes from: 
(1) the increased numbers, and wide $\lambda ^{2}$ range used for RM's immediately around the Galactic plane, and (2): our chosen smoothing scale, 
corresponding to $z \sim$ 1kpc at a distance of 6 kpc, preferentially detects features on a spiral arm and inter-arm scale up to $\sim 8$ kpc from the Sun. Our analysis is thus sensitive to the {\it large scale} underlying magnetic geometry of the Galactic disc and favours longer Galactic pathlengths. 

The RM variations in a band from $b = + 10^\circ$ to $-10^\circ$ about $l$ = 0$^\circ$ around the Galactic plane show clear and sharp sign changes. There is striking mirror symmetry about a reference direction $l = -5.5^\circ$. This gives a precisely determined, $l$ = 84.5 $\pm 1^\circ$, (i.e. $l = 90 - 5.5^\circ$) mean magnetic field direction in the vicinity of the Sun $-$  more precisely determined than in earlier analyses. In a 3-D representation this inward spiral tilt occurs over the approximate $\sim 1.5$ kpc $z$-height of the Milky Way disc, at least in our broad Galactic vicinity. We clearly rule out a circular magnetic field pattern near our Galactic radius. The distinct ``softening'' of the RM($l$) patterns at $|b|$ $\gtrsim 15^\circ$ in Figure 5 indicates that these global patterns are confined to $z$-heights of  ~1 - 2 kpc of the Milky Way's disc, and that other magnetic patterns emerge as we proceed further from the Galactic plane into the halo.     

We have shown in Section 3.2 that the prevailing magnetic field direction does not change sign across the galactic disc, i.e. that the magnetic field is even with respect $z=0$.  This applies at {\it all} $l$~ ranges at $|b| < 12^{\circ}$ that is, where $<{\bf B}>$ changes sign with $l$, the same sign is locally preserved on the opposite side of the Galactic plane. It is consistent with, but does not uniquely confirm, the quadrupole symmetry predicted by dynamo theory ({\it e.g.} Vall{\'e}e, 1992).

The improved definition of the magnetic structure of the Galactic disc is of interest for understanding, 
and comparing the global magnetic structure of the Milky Way and other galaxies, and 
their evolution ({\it e.g.} Beck {\it et al.} 1997, Shukurov {\it et al.} 2006), and for understanding the propagation of charged, 
very high energy cosmic ray nuclei through our Galaxy ({\it e.g.} The Pierre Auger Collaboration, 2007).


\section*{Acknowledgments}

The principal results of this paper form part of a Ph.D thesis completed in early 2009(K.N-M), and appeared, in part, in arXiv:0909.4753.
PPK acknowledges support from an NSERC (Canada), the U.S. DOE (LANL), the hospitality of CSIRO, and Prof. B. Gaensler at the U. of Sydney. 
PPK and KNM gratefully acknowledge support from Australian Research Council Federation Fellowships of 
Prof. Bryan Gaensler (FF0561298) and Prof. Ron Ekers (FF0345330) respectively. We are grateful to the anonymous 
referee for helpful comments.



\end{document}